\documentclass[twocolumn,aps,prl,showpacs,amsmath,
amssymb,superscriptaddress]{revtex4}

\usepackage{graphicx}
\usepackage{dcolumn}
\usepackage{bm}

\begin{document}

\title{``Spin-Flop'' Transition and Anisotropic
Magnetoresistance in Pr$_{1.3-x}$La$_{0.7}$Ce$_{x}$CuO$_4$: Unexpectedly
Strong Spin-Charge Coupling in Electron-Doped Cuprates}

\author{A. N. Lavrov}
\affiliation{Central Research Institute of Electric Power Industry,
Komae, Tokyo 201-8511, Japan}
\author{H. J. Kang}
\affiliation{Department of Physics and Astronomy, The University of
Tennessee, Knoxville, Tennessee 37996-1200, USA}
\affiliation{Condensed Matter Sciences Division, Oak Ridge National
Laboratory, Oak Ridge, Tennessee 37831-6393, USA}
\author{Y. Kurita}
\altaffiliation[Also at ]{Department of Physics, Tokyo
University of Science, Shinjuku-ku, Tokyo 162-8601, Japan.}
\author{T. Suzuki}
\altaffiliation[Also at ]{Department of Physics, Tokyo
University of Science, Shinjuku-ku, Tokyo 162-8601, Japan.}
\affiliation{Central Research Institute of Electric Power Industry,
Komae, Tokyo 201-8511, Japan}
\author{Seiki Komiya}
\affiliation{Central Research Institute of Electric Power Industry,
Komae, Tokyo 201-8511, Japan}
\author{J. W. Lynn}
\affiliation{NIST Center for Neutron Research, National Institute of
Standards and Technology, Gaithersburg, Maryland 20899, USA}
\author{S.-H. Lee}
\affiliation{NIST Center for Neutron Research, National Institute of
Standards and Technology, Gaithersburg, Maryland 20899, USA}
\author{Pengcheng Dai}
\email{daip@ornl.gov}
\affiliation{Department of Physics and Astronomy, The University of
Tennessee, Knoxville, Tennessee 37996-1200, USA}
\affiliation{Condensed Matter Sciences Division, Oak Ridge National
Laboratory, Oak Ridge, Tennessee 37831-6393, USA}
\author{Yoichi Ando}
\email{ando@criepi.denken.or.jp}
\affiliation{Central Research Institute of Electric Power Industry,
Komae, Tokyo 201-8511, Japan}

\date{\today}

\begin{abstract}

We use transport and neutron-scattering measurements to show that a
magnetic-field-induced transition from noncollinear to collinear spin
arrangement in adjacent CuO$_2$ planes of lightly electron-doped
Pr$_{1.3-x}$La$_{0.7}$Ce$_{x}$CuO$_4$ ($x=0.01$) crystals affects
significantly both the in-plane and out-of-plane resistivity. In the
high-field collinear state, the magnetoresistance (MR) does not saturate,
but exhibits an intriguing four-fold-symmetric angular dependence,
oscillating from being positive at ${\bf B}$$\,\parallel\,$[100] to being
negative at ${\bf B}$$\,\parallel\,$[110]. The observed MR of more than
30\% at low temperatures induced by a modest modification of the spin
structure indicates an unexpectedly strong spin-charge coupling in
electron-doped cuprates.

\end{abstract}


\maketitle

High-$T_c$ superconductivity (SC) in cuprates emerges as the parent
antiferromagnetic (AF) insulator is doped with charge carriers, either
holes or electrons. Despite this apparent symmetry with respect to doping,
it still remains unclear whether the mechanism of SC in both cases is the
same. It is generally believed that in the hole-doped cuprates, the SC
pairing originates from an interplay between the doped holes and AF spin
correlations. Indeed, many observations, including a fast suppression of
the N\'{e}el order by doped holes \cite{LSCO_TN} which results in the
``spin-glass'' state \cite{LSCO_TN, Clast_SG, suscept}, and a strong
tendency to form spin-charge textures -- ``stripes'' \cite{NdSr}, point to
a very strong coupling between the charge and spin degrees of freedom.

The behavior of doped {\it electrons} looks, however, much different.
Electron doping suppresses the AF order at virtually the same slow rate as
the substitution of magnetic Cu$^{2+}$ ions with non-magnetic Zn$^{2+}$
\cite{elect_Zn,TN_PCCO}, and does not induce any incommensurability in the
spin correlations \cite{commens}. This has been taken as evidence that the
electrons merely dilute the spin system \cite{LSCO_TN, elect_Zn,TN_PCCO}.
Apparently, if the charge transport and spin correlations are actually
decoupled in the electron-doped cuprates, the SC pairing should have a
non-magnetic origin as well. A recent discovery of the magnetic-field
induced AF order in superconducting Nd$_{2-x}$Ce$_x$CuO$_4$
\cite{NCCO_AF,masato} has shown, however, that antiferromagnetism and
superconductivity may be closely related in these compounds.

To probe the spin-charge coupling, one can determine how the charge
transport responds to such relatively weak changes in the spin structure
as spin-flop or metamagnetic transitions. In hole-doped
La$_{2-x}$Sr$_x$CuO$_4$, for instance, the conductivity changes by up to
several times \cite{MR_WF, LSCO_MR}. In this Letter, we use neutron
scattering and magnetoresistance (MR) measurements to study the effect of
magnetic field on the spin structure and anisotropic conductivity of
lightly electron-doped Pr$_{1.3-x}$La$_{0.7}$Ce$_{x}$CuO$_4$ (PLCCO)
single crystals. We find that both the in-plane and out-of-plane
resistivity ($\rho_{ab}$ and $\rho_{c}$) are surprisingly sensitive to
spin reorientation, with $\Delta\rho_{ab}/\rho_{ab}$ exceeding 30\% at low
temperatures -- the same scale as in hole-doped La$_{2-x}$Sr$_x$CuO$_4$
\cite{LSCO_MR}. This result indicates that in electron-doped cuprates the
charge transport exhibits a similar degree of coupling to magnetism as in
the hole-doped ones, and therefore the superconductivity in both systems
may have a universal origin.

High-quality PLCCO single crystals with $x=0.01$ (mosaicity $<1^\circ$)
were grown by the traveling-solvent floating-zone technique and annealed
at $\approx860^\circ$C in pure argon to remove excess oxygen. The partial
substitution of Pr with La was used to stabilize the crystal growth,
without introducing significant lattice distortions \cite{PLCCO}. Neutron
scattering measurements were performed on the BT-2 and SPINS triple-axis
spectrometers at the NIST Center for Neutron Research. We label
wavevectors ${\bf Q}=(q_x,q_y,q_z)$ in \AA$^{-1}$ as
$(H,K,L)=(q_xa/2\pi,q_ya/2\pi,q_zc/2\pi)$ in the reciprocal lattice units
(r.l.u.) suitable for the tetragonal unit cell of PLCCO (space group
$I4/mmm$, $a=3.964$ and $c=12.28$ \AA \ are in-plane and out-of-plane
lattice paramters, respectively). In this notation, [100]/[010] and
[110]/[$\overline{1}$10] are along the Cu-O-Cu bond direction and the
diagonal Cu-Cu direction, respectively. The experimental details are
described in Refs. \cite{NCCO_AF,masato}.

Resistivity measurements were carried out by the ac four-probe method on
the {\it same} crystal used for neutron measurements. It was cut and
polished into suitable shapes: $3.1\times1\times0.45$ mm$^3$ for
$\rho_{ab}$ and $\approx1\times1\times1$ mm$^3$ for $\rho_c$. The MR was
measured by sweeping the magnetic field between $\pm 14$ T at fixed
temperatures stabilized by a capacitance sensor with an accuracy of $\sim
1$ mK.

\begin{figure}[!t]
\includegraphics*[width=7.2cm]{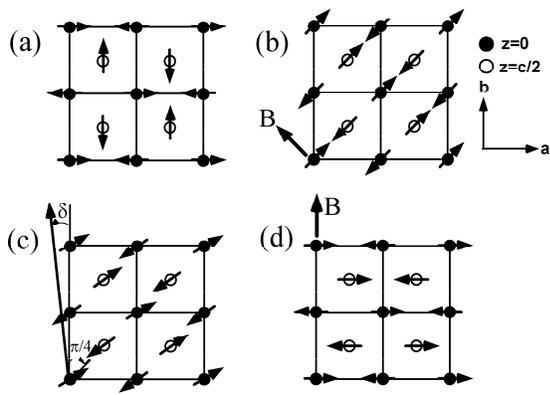}
\caption{Field-induced transition from noncollinear to col\-linear spin
arrangement in Pr$_2$CuO$_4$. (a) Zero-filed noncollinear spin structure;
only Cu spins are shown. (b) - (c) Collinear spin-flop states induced by
(b) a magnetic field applied along the Cu-Cu direction; (c) a magnetic
field tilted from [010]; and (d) ${\bf B}$$\,\parallel\,$[010].}
\label{fig1}
\end{figure}

\begin{figure}[!b]
\includegraphics*[width=8.1cm]{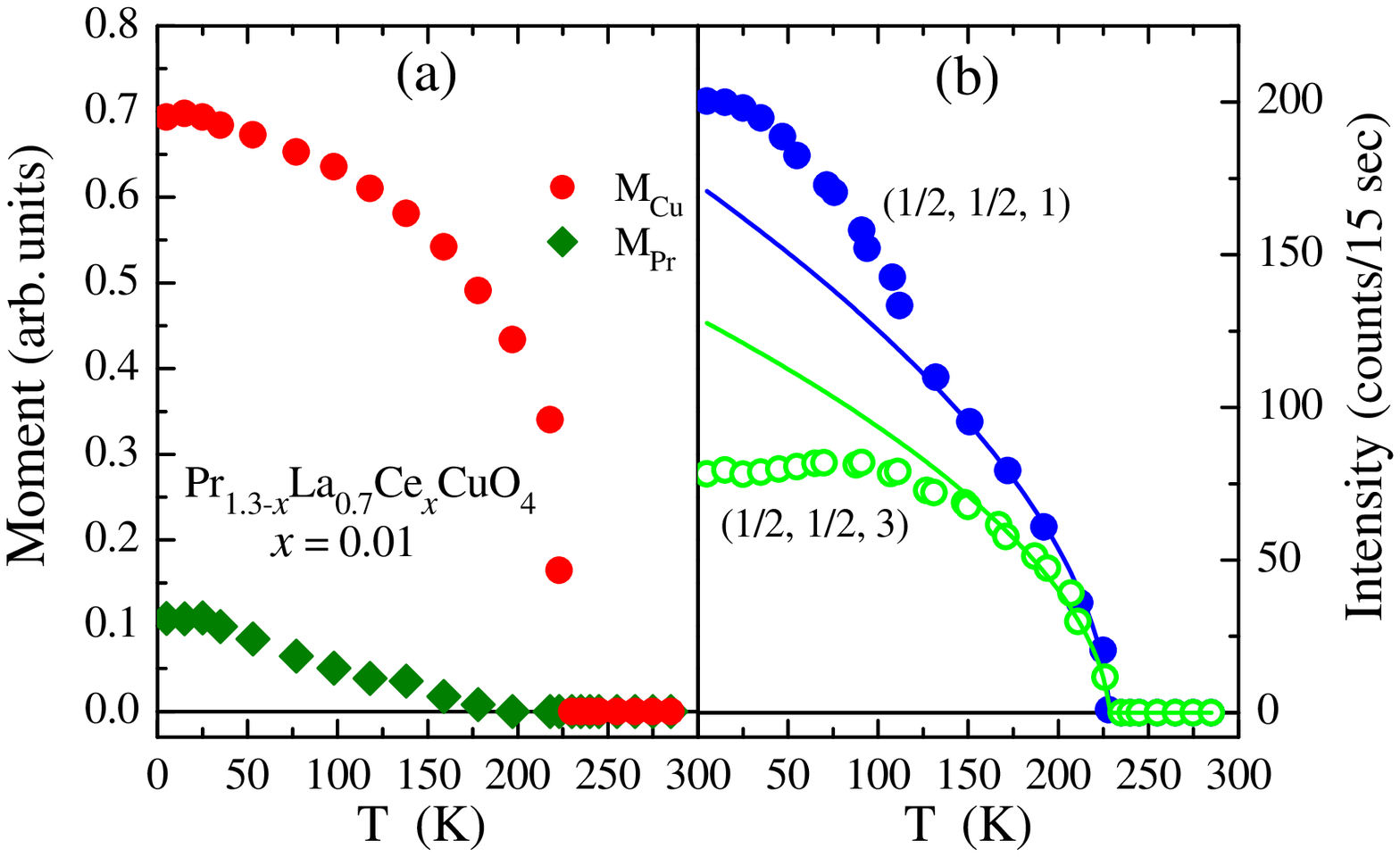}
\caption{(a) Temperature dependence of the Cu$^{2+}$ and Pr$^{3+}$
moments in PLCCO ($x=0.01$). (b) Integrated intensity of the
($\frac{1}{2},\frac{1}{2},1$) and ($\frac{1}{2},\frac{1}{2},3$) magnetic
peaks. The ordered moments are estimated by normalizing the magnetic
intensity to the weak (1,1,0) nuclear Bragg peak without considering the
absorption and extinction effects \cite{masato}. The solid
lines are power-law fits describing the contribution of Cu spins
\cite{Sumarlin}.}
\label{fig2}
\end{figure}

The peculiar spin structure of Pr$_2$CuO$_4$ (PCO) is interesting in its
own right. While a strong intraplane exchange drives the AF spin ordering
within CuO$_2$ planes, all the isotropic exchange interactions {\it
between} the planes are perfectly canceled out due to the body-centered
tetragonal crystal symmetry. The three-dimensional ordering [Fig. 1(a)]
that sets in below the N\'{e}el temperature $T_N= 250-285$ K
\cite{Sumarlin, PseudoD2, PCO_QT} is governed by weak pseudodipolar (PD)
interactions, which favor a noncollinear orientation of spins in adjacent
planes (alternating along the [100] and $[0\overline{1}0]$ directions)
\cite{Sumarlin, PseudoD1, PseudoD2, PCO_QT}. A unique feature of the
interplane PD interaction is that its energy does not change if the spin
sublattices of adjacent CuO$_2$ planes rotate in {\it opposite} directions
\cite{PseudoD1, PseudoD2, PCO_QT}. Such a continuous spin rotation can be
induced by a magnetic field parallel to Cu-Cu direction, which easily
converts the noncollinear structure of Fig. 1(a) into a collinear one with
spins along the [110] direction [Fig. 1(b)]. Note that while these
diagonal directions are hard spin axes in the non-collinear phase, they
become the easy axes in the collinear one. A perfectly aligned field ${\bf
B}$$\,\parallel\,$[010] causes a first-order transition directly to the
spin-flop phase [Fig. 1(d)], while at intermediate field directions the
magnetic field first induces a transition into the collinear phase [Fig.
1(c)], and then smoothly rotates the spins to align them perpendicular to
the field \cite{PCO_QT}.

\begin{figure}[!tb]
\includegraphics*[width=8.2cm]{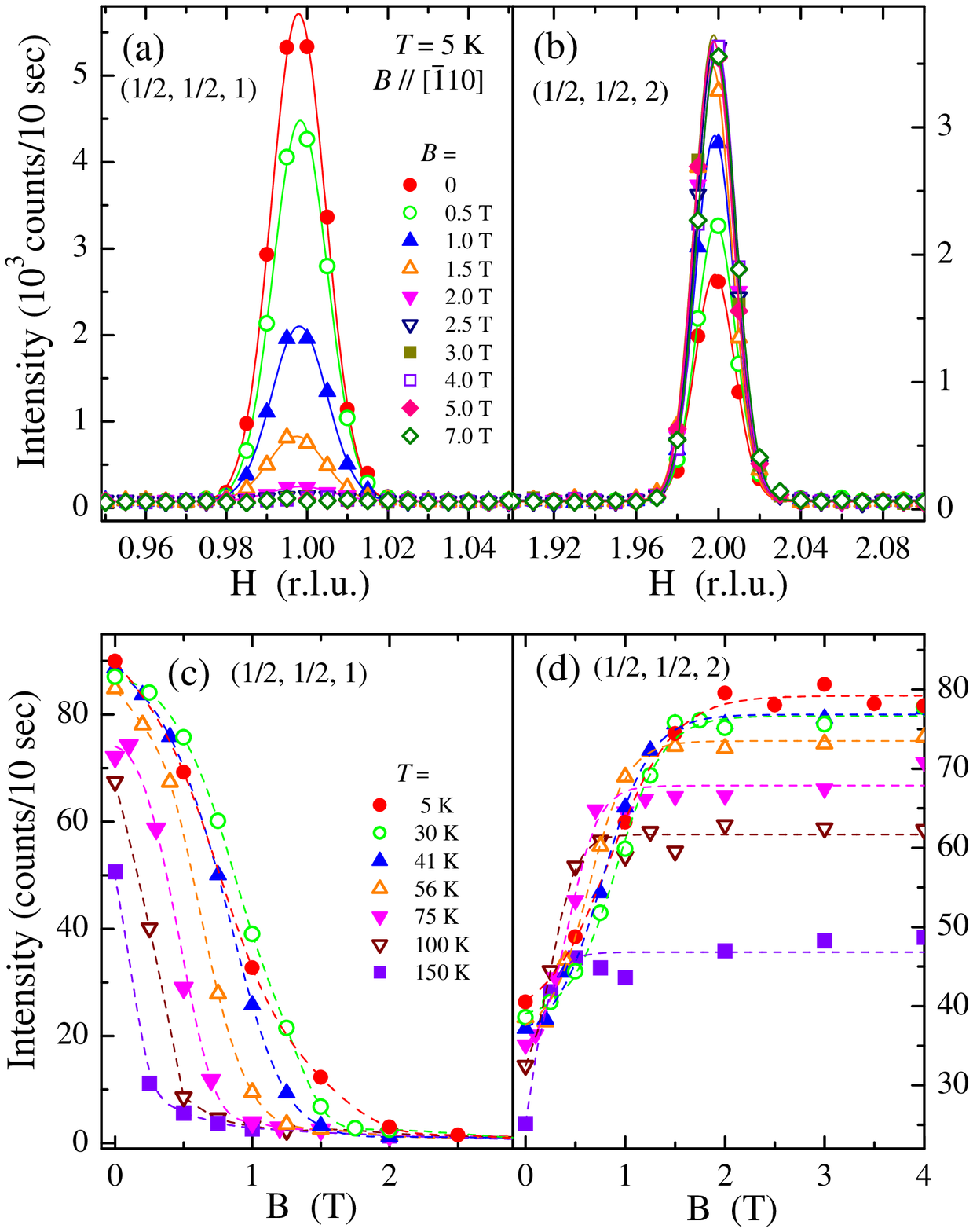}
\caption{(a), (b) Effect of the ${\bf B}$$\,\parallel\,$$[\overline{1}10]$
field on ($\frac{1}{2},\frac{1}{2},1$) and ($\frac{1}{2},\frac{1}{2},2$)
magnetic peaks at 5 K. (c), (d) Field dependence of the integrated
intensity at various temperatures. We note that the critical field for
spin-flop transition in PLCCO is lower than that of PCO \cite{Sumarlin}.}
\label{fig3}
\end{figure}

\begin{figure*}[!t]
\includegraphics*[width=16.9cm]{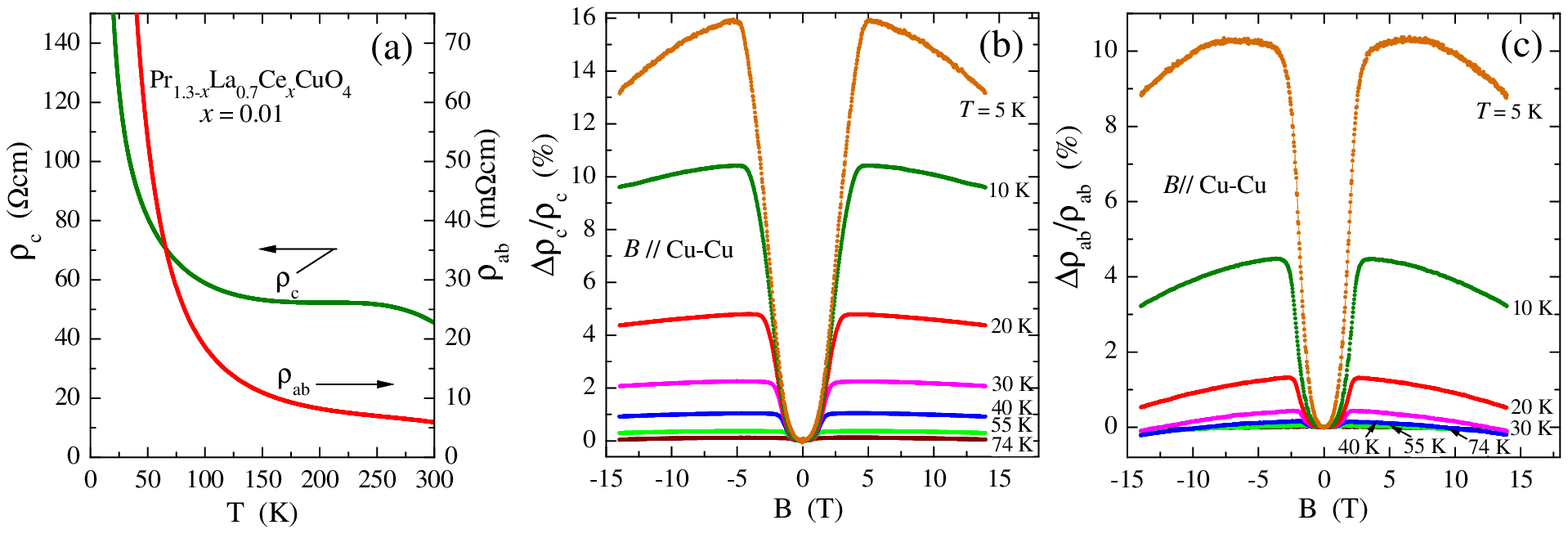}
\caption{(a) In-plane and out-of-plane resistivity of PLCCO ($x=0.01$)
single crystals. The MR in $\rho_{c}$ (b) and $\rho_{ab}$ (c) measured for
the in-plane magnetic field ${\bf B}$$\,\parallel\,$$[\overline{1}10]$.}
\label{fig4}
\end{figure*}

The neutron diffraction measurements at zero field on the $(1/2,1/2,L)$
magnetic Bragg peaks ($L=0,1,2,3,4$) show that in our PLCCO ($x=0.01$) the
Cu$^{2+}$ spins order into the same non-collinear structure as in pure
PCO, albeit at a somewhat lower $T_N\approx 229$ K (Fig. 2). The reduced
$T_N$ is probably due to a partial substitution of Pr$^{3+}$ with
non-magnetic La$^{3+}$, as well as to doped electrons. Similar to PCO
\cite{Sumarlin}, the Pr$^{3+}$ ions in PLCCO can be polarized by the
ordered Cu$^{2+}$ moment. Upon cooling below 100-150 K, the exchange field
of the Cu$^{2+}$ spins induces a small (up to $\sim 0.1$ $\mu_{B}$)
ordered moment on the Pr$^{3+}$ ions (Fig. 2).

Figure 3 shows the effect of a ${\bf B}$$\,\parallel\,$$[\overline{1}10]$
field on the $(1/2,1/2,1)$ and $(1/2,1/2,2)$ magnetic peaks at various
temperatures. Upon increasing the magnetic field, the peak intensity
changes, indicating a continuous non-collinear to collinear phase
transition. Indeed, for the collinear spin arrangement [Fig. 1(b)], the
magnetic intensity vanishes at $(1/2,1/2,L)$ with $L=1,3,5$. As can be
seen in Figs. 3(c) and 3(d), the critical field for the non-collinear to
collinear (``spin-flop'') transition, $B_c$, increases from less than 0.5
T at 150 K to $\sim$2 T at 5 K. In comparison, the first-order spin-flop
transition for ${\bf B}$$\,\parallel\,$[010] was reported to take place at
several time larger fields \cite{PCO_QT} and a $c$-axis aligned field does
not change the noncollinear spin structure \cite{masato}.

\begin{figure*}[!ht]
\includegraphics*[width=15.6cm]{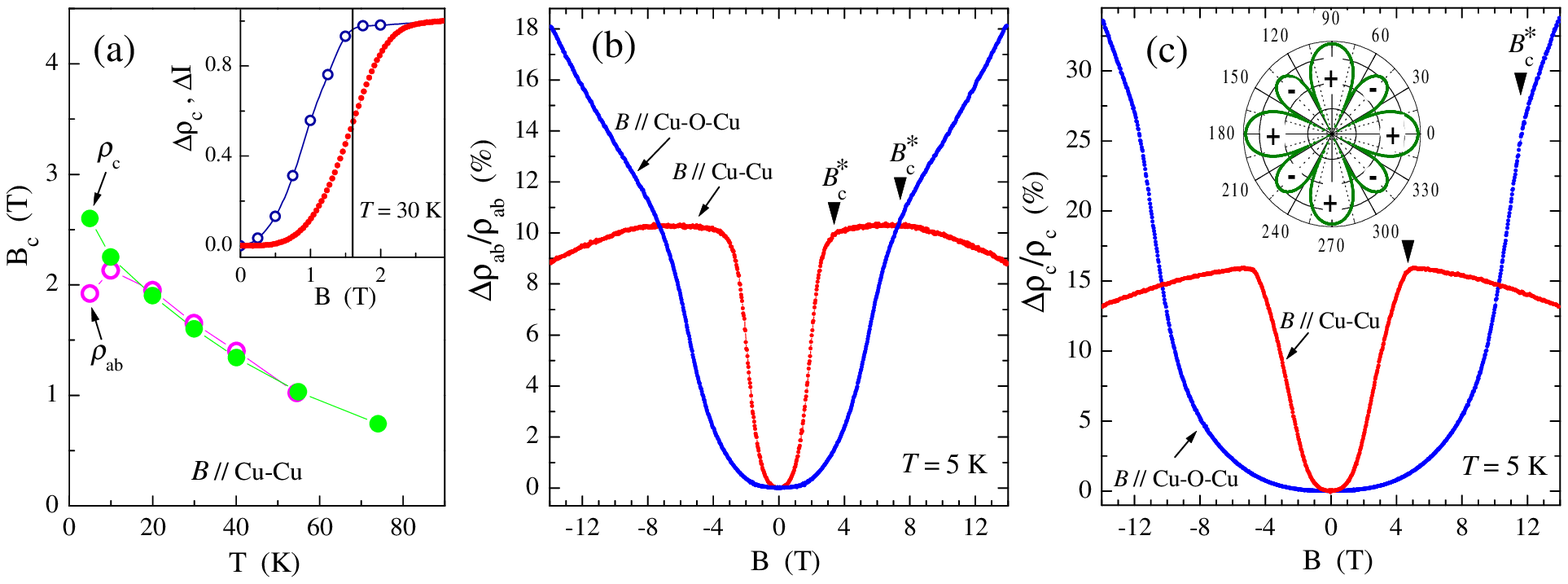}
\caption{(a) The critical field $B_c$ determined from peaks in
$d\rho_{ab}/dB$ and $d\rho_{c}/dB$ for ${\bf
B}$$\,\parallel\,$$[\overline{1}10]$. In the inset to (a), the normalized
field dependence of $\rho_c$ ($\bullet$) is compared with that of the
($\frac{1}{2},\frac{1}{2},1$)-peak intensity ($\circ$).
$\Delta\rho_{ab}/\rho_{ab}$ (b) and $\Delta\rho_{c}/\rho_{c}$ (c) for two
directions of the in-plane magnetic field. The angular dependence of the
high-field MR is sketched in the inset to (c).}
\label{fig5}
\end{figure*}

The transport properties of lightly electron-doped PLCCO differ from those
of its hole-doped analog LSCO or YBa$_2$Cu$_3$O$_{6+x}$ (YBCO). In
contrast to hole-doped cuprates \cite{LSCO_MR,mobility,YBCO}, the doping
of 1\% of electrons into the CuO$_2$ planes appears to be insufficient to
induce metallic in-plane conduction in PLCCO, and both $\rho_{ab}$ and
$\rho_{c}$ grow upon cooling below room temperature [Fig. 4(a)]. It is
worth noting also that lightly doped PLCCO turns out to be one of the most
anisotropic cuprates with $\rho_{c}/\rho_{ab}\sim8000$ at room temperature
-- an order of magnitude larger than in LSCO and YBCO \cite{LSCO_MR,YBCO}.

In further contrast to hole-doped cuprates \cite{LSCO_MR,YBCO}, no anomaly
is detected at the N\'{e}el transition in PLCCO either in the in-plane or
out-of-plane resistivity. At a first glance, this supports the view that
the charge motion in electron-doped PLCCO is virtually decoupled from spin
correlations, and one therefore would expect the conductivity to ignore
the spin reorientation sketched in Fig. 1. Surprisingly, the experiment
shows that this is not the case, and instead of being field-independent,
both $\rho_{ab}$ and $\rho_{c}$ exhibit a considerable increase upon
transition into the collinear state [Figs. 4(b) and 4(c)]. We have
confirmed that this MR is of the spin origin and contains no orbital
terms, since no difference was observed in $\Delta\rho_{ab}/\rho_{ab}$ for
fields applied parallel or perpendicular to the current. Moreover,
$\Delta\rho_{ab}/\rho_{ab}$ and $\Delta\rho_{c}/\rho_{c}$ demonstrate a
remarkable similarity both in magnitude and in field dependence, in spite
of the huge resistivity anisotropy. Finally, no MR anomaly is observed
when a $c$-axis aligned field is applied, consistent with the absence of a
spin-flop transition for such field orientation \cite{masato}.

The MR behavior in Fig. 4 is clearly reminiscent of that in LSCO
\cite{LSCO_MR}, though there are two important differences. First is the
sign of the anomalous MR, which is always positive in PLCCO, but negative
in LSCO. Second, the MR features in LSCO and YBCO become discernible as
soon as the AF order is established, but in PLCCO they appear at
temperatures much lower that $T_N$ (at $T<70-100$ K), and quickly gain
strength upon decreasing temperature (Fig. 4). The latter indicates that
some other factors, such as magnetic moments of Pr$^{3+}$ or a structural
instability \cite{Ortho}, that come into play at low temperature, may be
relevant to the observed MR.

A comparison of the neutron and resistivity data reveals one more
interesting feature, namely, the transitions observed by these two probes
do not match each other [inset to Fig. 5(a)]. One can see that the charge
transport ignores the initial spin rotation, and the steepest resistivity
variation is observed at $B_c$, where the collinear structure is
established. Although $B_c$ changes substantially with temperature [Fig.
5(a)], the apparent shift in the transitions holds consistently, with the
peak in $d\rho/dB$ roughly coinciding with the end of the transition
observed by neutron scattering.

As the magnetic field deviates from the Cu-Cu direction [Fig. 1(c)], the
spin-flop transition shifts towards higher fields, reaching ultimately
$B_c\sim 12$ T for $B$$\,\parallel\,$[010]; the MR behavior for these two
field orientations is compared in Figs. 5(b) and 5(c) \cite{align}. It
becomes immediately clear from these figures that the step-like increase
of the resistivity upon the transition to the collinear state does not
make a complete story. Regardless of the field direction within the $ab$
plane, the resistivity exhibits roughly the same increase at the spin-flop
transition, but then (at $B>B_c^*$) it keeps changing without any sign of
saturation [Figs. 5(b) and 5(c)]. Even more surprising is that this
high-field MR changes its sign depending on the field direction, as is
schematically drawn in the inset to Fig. 5(c). One can conceive a spin
structure upon rotating the high magnetic field within the $ab$ plane in
the following way: the spins always keep the collinear arrangement and
rotate as a whole, being almost perpendicular to the magnetic field (Fig.
1). Our data show that the resistivity goes down as the spin direction
approaches one of the two equivalent spin easy axes (Cu-Cu directions) and
increases at the spin hard axes (Cu-O-Cu directions) [inset to Fig. 5(c)].
Note that the resistivity changes are rather large,
$\Delta\rho_{ab}/\rho_{ab}$ reaches $\approx 18$\% at $T=5$ K and exceeds
32\% at 2.5 K, indicating that the magnetic field ${\bf
B}$$\,\parallel\,$[100] can effectively localize the doped electrons.

Apparently, the fascinating MR oscillations in Fig. 5 cannot originate
from simple ``spin-valve'' effects, since at high fields the spin
structure always stays collinear, and all that changes is the relative
orientation of spins with respect to the crystal axes. The MR may be
related to 2D spin fluctuations that were found to survive far above
$B_c$, as manifested in the diffuse neutron scattering \cite{petitgrand1},
or to some unusual coupling of the charge transport with low-energy spin
dynamics. Though the exact mechanism of the revealed MR features still
remains to be understood, what is certain is that the charge carriers in
electron-doped cuprates appear to have a remarkably strong coupling with
the spin order, which should play an important role in determining their
physical properties.

Upon preparing this paper, we became aware of similar MR features observed
for Pr$_{1.85}$Ce$_{0.15}$CuO$_4$ \cite{Fournier}, which gives evidence
that the strong spin-charge coupling survives up to much higher
electron-doping levels, that are relevant for the superconducting state.

\begin{acknowledgments}
We thank K. Segawa and Shiliang Li for technical assistance.
This work was in part supported
by the US NSF DMR-0139882
and DOE under contract No. DE-AC-00OR22725 with UT/Battelle, LLC.

\end{acknowledgments}


\begin{thebibliography}{99}
\bibitem{align} For the case of $B$$\,\parallel\,$[010], the transition
field is extremely sensitive to the crystal alignment; just a few degrees
tilting of the $\rho_{ab}$ sample in Fig. 5(b) notably shifted the
transition to low fields, as compared with the better aligned $\rho_{c}$
sample in Fig. 5(c).
\end{thebibliography}
\end{document}